      [1999/12/01 v1.4c Il Nuovo Cimento]
\documentclass{cimento}

\def\farcs{\hbox{$\,.\!\!^{''}$}}
\def\fs{\hbox{$\,.\!\!^{s}$}}

\usepackage{graphicx} 
\title{GRB990123: multiwavelength afterglow study}

\author{E.~Maiorano\from{ins:a}\from{ins:b}\thanks{e-mail: {\tt 
maiorano@bo.iasf.cnr.it}}\ETC,
N.~Masetti\from{ins:a},
E.~Palazzi\from{ins:a},
F.~Frontera\from{ins:a}\from{ins:c}, 
P.~Grandi\from{ins:a},
E.~Pian\from{ins:d},
L.~Amati\from{ins:a},
L.~Nicastro\from{ins:e},
P.~Soffitta\from{ins:f},
A.~Corsi\from{ins:f}, 
L.~Piro\from{ins:f},
L.A.~Antonelli\from{ins:g},
E.~Costa\from{ins:f},
M.~Feroci\from{ins:f},
J.~Heise\from{ins:h}
\atque \\
J.J.M.~in 't Zand\from{ins:h}
}

\instlist{
\inst{ins:a} Istituto di Astrofisica Spaziale e Fisica Cosmica -- (INAF) 
sez. di Bologna, Italy
\inst{ins:b} Dipartimento di Astronomia, Universit\`a di Bologna, Italy
\inst{ins:c} Dipartimento di Fisica, Universit\`a di Ferrara, Italy
\inst{ins:d} Osservatorio Astronomico di Trieste-- (INAF), Italy
\inst{ins:e} Istituto di Astrofisica Spaziale e Fisica Cosmica -- (INAF) 
sez. di Palermo, Italy
\inst{ins:f} Istituto di Astrofisica Spaziale e Fisica Cosmica -- (INAF) 
sez. di Roma, Italy
\inst{ins:g} Osservatorio Astronomico di Roma -- (INAF) Monte Porzio 
Catone, Italy
\inst{ins:h} SRON, Sorbonnelaan 2, NL-3584 CA Utrecht, The Netherlands
}

\PACSes{
\PACSit{95.75.Fg}{Spectroscopy and spectrophotometry}
\PACSit{95.85.Nv}{X--ray}
\PACSit{98.70.Rz}{gamma--ray sources; gamma--ray bursts}
}

\begin{document}

\maketitle

\vspace{-0.5cm}

\begin{abstract} 
We report on the {\it BeppoSAX} and multiwavelength data analysis of the
afterglow of Gamma-Ray Burst (GRB) 990123. Mainly due to its exceptional
brightness, this is the only source for which the Wide Field Cameras have
allowed an early detection of the X--ray afterglow between $\sim$20 and 60
min after the GRB trigger. Besides, again for the first time, high-energy
emission from the afterglow was detected up to 60 keV. The backwards
extrapolation of the afterglow decay smoothly reconnects with the late GRB
emission, thus indicating that both are consistent with being produced by
the same phenomenon. For the X--ray afterglow we found a power--law decay
with index $\alpha_{\rm X} = 1.46\pm0.04$; the spectrum has a power-law
shape with photon index $\Gamma \sim 2$. An extensive set of
multiwavelength observations on the afterglow, collected from the
literature and made during the {\it BeppoSAX} pointing, allowed
constructing a Spectral Flux Distribution. We performed an analysis of the
GRB afterglow emission in the context of the ``fireball'' model.
\end{abstract}

\vspace{-1.cm}
\section{Introduction}
\vspace{-.2cm}

The Gamma-Ray Burst (GRB) 990123 triggered the {\it
BeppoSAX}~\cite{ref:boe} Gamma Ray Burst Monitor (GRBM) on 1999 January
23.4078 UT~\cite{ref:pir} and was simultaneously detected near the center
of the field of view in Wide Field Camera (WFC) no. 1~\cite{ref:jag}, with
a localization uncertainty of 2' (error circle radius) at coordinates
(J2000) RA = 15$^{\rm h}$ 25$^{\rm m}$ 29$\fs$00, DEC = +44$^{\circ}$
45$'$ 00$\farcs$5~\cite{ref:fer}. The prompt emission~\cite{ref:cor}
lasted about 100 s in 40--700 keV and showed two major peaks (Fig.  1,
left panel). At the lower (2--10 keV) energies in the WFC data,
atmospheric absorption played an important role, as it affected
substantially the soft X--ray measurements close to the end of the
observation. In this burst for the first time the optical flash was
observed by the Robotic Optical Transient Search Experiment (ROTSE), the
OT peaked at magnitude $V\sim$ 9 about 50 s after the GRB onset. The
afterglow optical spectrum showed an absorbtion system at redshift
$z$ = 1.600~\cite{ref:kul,ref:and}.

Here we present the data on the X--ray and multiwavelength afterglow of
GRB990123. A more extensive treatment of the data presented here can be
found in Maiorano et al.~\cite{ref:maio}. When not otherwise indicated,
errors for X--ray spectral parameters will be reported at 90\% confidence
level ($\Delta\chi^2$ = 2.7 for one parameter fit), while errors for other
parameters will be at 1$\sigma$; upper limits will be given at 3$\sigma$.


\begin{figure}
\vspace{-.45cm}
\hspace{-.5cm}
\includegraphics[height=8.3cm]{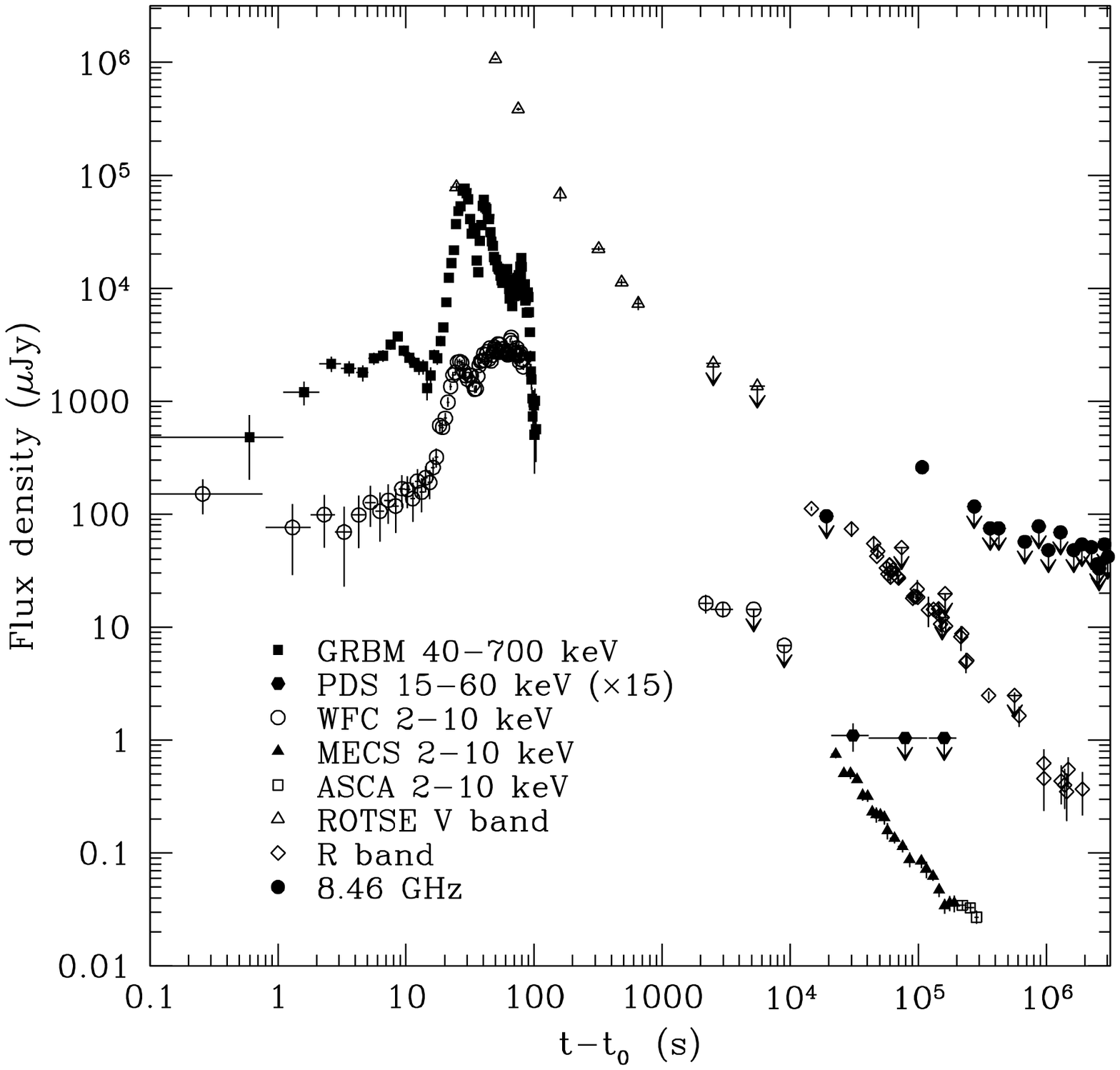}
\hspace{-.5cm}
\includegraphics[height=8.0cm]{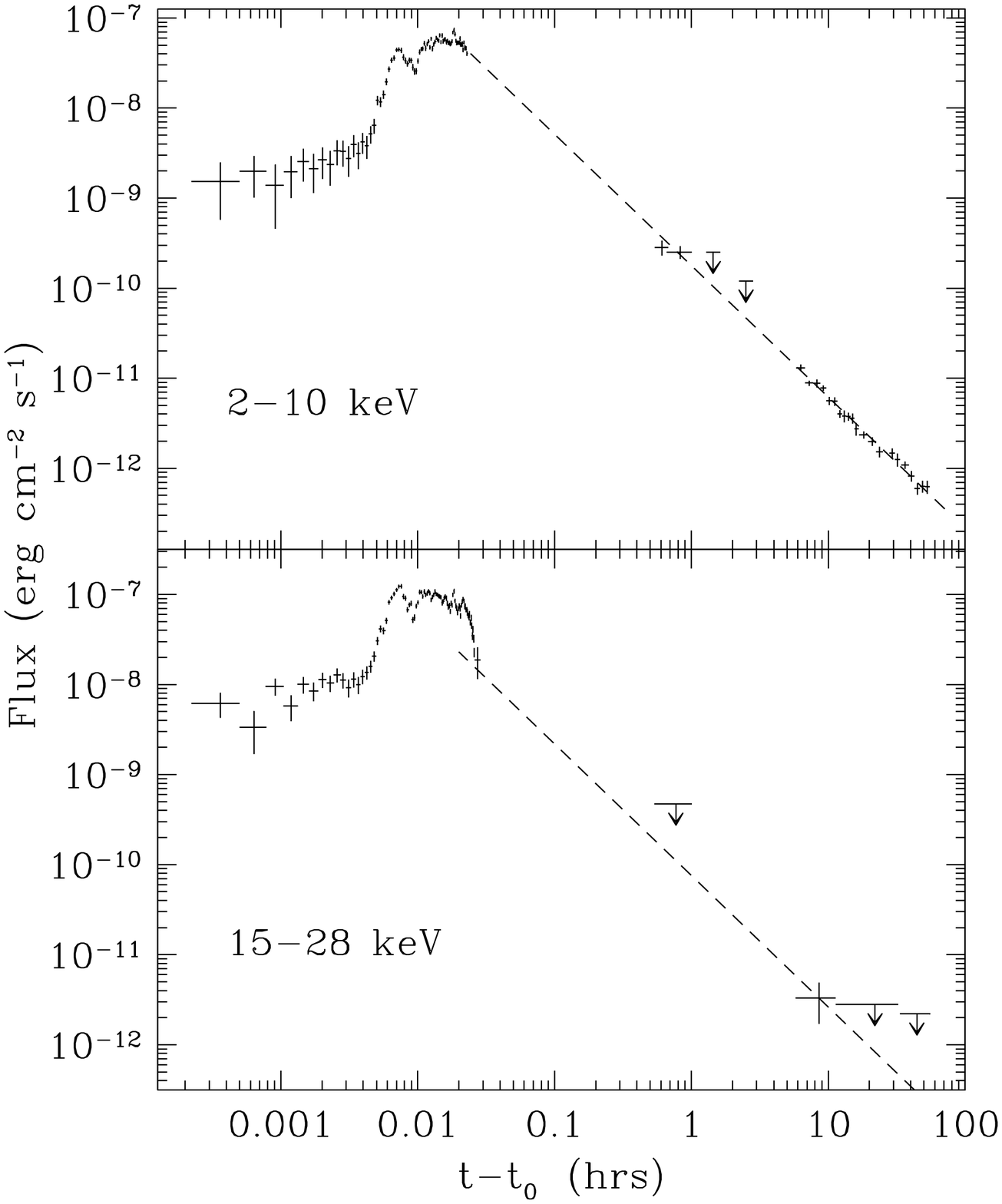} 
\vspace{-.9cm}
\caption{Left panel: multiwavelength light curves; $t_0$ 
corresponds to the time of the GRB onset. Right panel: 2--10 keV and 
15--28 keV light curves from the prompt event to the afterglow. The dashed 
line is the best-fit decay obtained fron the X--ray afterglow data.} 
\end{figure}

\vspace{-.3cm}
\section{Observations and data analysis}
\vspace{-.2cm}

Two {\it BeppoSAX} Narrow-Field Instruments (NFI) observations were
carried out in Target of Opportunity (ToO) mode. The observations started
$\sim$ 6 hr after the trigger and continued during the two following days
(see Table 1). A fading X--ray source inside the WFC error box was
detected almost at the center of the LECS and MECS detectors and it was
identified as the X-ray afterglow of GRB990123~\cite{ref:pir1}. Afterglow
spectra and light curves for the NFI data were then accumulated and
extracted.

\vspace{-.3cm}
\section{Results}
\vspace{-.2cm}

\subsection{Light curves}

In Fig. 1 (left panel) $\gamma$--ray (40--700 keV), X--ray (2--10 keV),
optical $V$--band~\cite{ref:ake}, optical
$R$--band~\cite{ref:cat,ref:gal,ref:kul,ref:fru}, and radio (36
mm,~\cite{ref:kul}) light curves of the prompt burst and afterglow of
GRB990123 are shown. Time is given in seconds since the GRBM trigger. A
detailed treatment of GRBM and WFC data can be found in Corsi et
al.~\cite{ref:cor}. The {\it ASCA} data points acquired after the end of
the {\it BeppoSAX} ToOs are from Yonetoku et al.~\cite{ref:yon}. The host
galaxy~\cite{ref:blo} was subtracted from the $R$ data. Gunn--$r$
magnitudes were converted into $R$ band following Fruchter et
al.~\cite{ref:fru}. $V$ and $R$ data were corrected for the Galactic
foreground reddening assuming $E(B-V) = 0.016$~\cite{ref:sch}. Using a
power--law model to describe the temporal decay, the 2--10 keV MECS
measurements are well fit with an index $\alpha_{\rm X} = 1.46\pm0.04$
(Fig. 1, right panel, where the dashed line indicates the best--fit decay
of the X--ray afterglow data). We obtained a 2--10 keV light curve from
the prompt event to the afterglow, using WFC and MECS data. Moreover, as
it is the first time in which an X--ray afterglow is detected above 10
keV, we derived the light curve in the 15--28 keV energy range using WFC
data for the prompt emission and PDS data for the afterglow (Fig. 1, right
panel). The backwards extrapolation of the 2--10 keV power--law smoothly
reconnects with the late--time WFC data points and upper limits,
suggesting that the last WFC points already represent the afterglow
emission. Thus, this is the only source for which the Wide Field Cameras
have allowed an early detection of the X--ray afterglow between $\sim$20
and 60 min after the GRB trigger.

\vspace{-.2cm}
\subsection{Spectra}


\begin{figure}
\vspace{1.7cm}
\mbox{\includegraphics[height=7.0cm,angle=-90]{maiorano_f2l.ps}}
\quad

\vspace{-6.5cm}
\hspace{7cm}
\includegraphics[width=7.0cm]{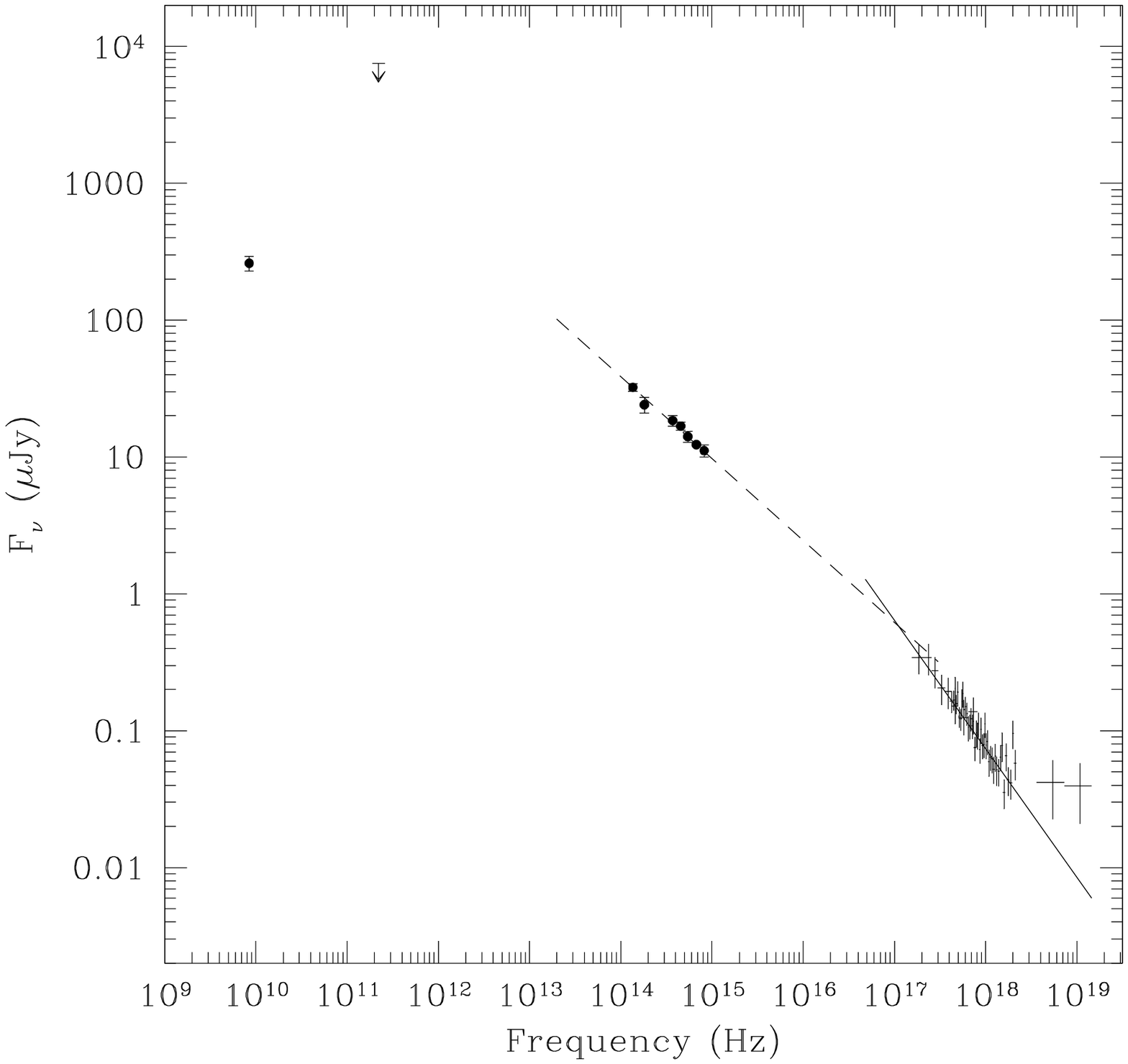}
\vspace{-1cm}
\caption{Left panel: the 0.6--60 keV afterglow during the first 20 ks of
the {\it BeppoSAX} ToO1. Right panel: SFD at 24.65 Jan. 1999 UT. The
optical, NIR and radio data are from Galama et al.~\cite{ref:gal}. As
regards the radio data only those acquired strictly within 0.03 d of the
reference epoch above are included in the plot. The dashed line is the
best--fit power--law ($\beta_{\rm opt} = 0.60$) describing optical and NIR
data, the solid line is the power--law ($\beta_{\rm X} = 0.94$) which best
fits the {\it BeppoSAX} NFI data.}
\end{figure}

At the beginning of the first ToO the X--ray (0.6--60 keV) afterglow
spectrum has a very good statistical quality and for the first time a GRB
X--ray afterglow was detected up to 60 keV. A simultaneous combined LECS,
MECS and PDS spectrum was obtained in the first 20 ks of the observation,
we fitted this spectrum using a photoelectrically absorbed power--law with
a photon index $\Gamma = 1.94^{+0.12}_{-0.13}$. As the value of $N_{\rm
H}$ obtained in this way ($0.9^{+15}_{-0.9} \times 10^{20}$ cm$^{-2}$), is
consistent with the Galactic column density along the GRB direction, we
fixed its value to the Galactic one ($1.98 \times 10^{20}$
cm$^{-2}$)~\cite{ref:dic}. The spectra accumulated over the following two
time intervals (76.2 ks of ToO1 and the whole ToO2, where the afterglow
was detected in the 0.6--10 keV range only) were also analyzed, and the
fits were made assuming again a power--law description; the best--fit
photon index values were found to be completely consistent with that
obtained in the first 20 ks spectrum (see Table 1). Thus we can say
that the 3 time resolved spectra are consistent with no spectral
variation, i.e., the X--ray afterglow evolution is achromatic.


\begin{table}
\caption{Results of the time-resolved spectral fits of the afterglow {\it 
BeppoSAX} NFI observations. In each case the N$_{\rm H}$ amount (in 
squared parentheses) was fixed at the Galactic value. Errors are given at 
90\% confidence level.}
\vspace{-.5cm}
\begin{center}
\begin{tabular}{cccccc}
\noalign{\smallskip}
\hline
\noalign{\smallskip}
ToO & Start time & duration & $\Gamma$ & N$_H$ & $\chi^2$/dof  \\
 & (Jan 1999 UT) & (ks) &  & (10$^{20}$ cm$^{-2}$) &  \\
\hline
\noalign{\smallskip}
1 (1$^{\rm st}$ part) & 23.6495 & 20 & $1.94^{+0.12}_{-0.13}$ & [1.98] & 
40/50  \\
1 (2$^{\rm nd}$ part) & 23.8810 & 76.2 & $2.07^{+0.11}_{-0.12}$ & [1.98] & 
56/54  \\
2 & 24.8132 & 76.5 & $1.86^{+0.29}_{-0.29}$ & [1.98] & 16/16  \\
\noalign{\smallskip}
\hline
\end{tabular}
\end{center}
\end{table}

\vspace{-.2cm}
\subsection{Spectral Flux Distribution}

We considered the Spectral Flux Distribution (SFD) already presented by
Galama et al.~\cite{ref:gal} and we completed it with the NFI X--ray data.
We referred all data points to date 24.65 Jan 1999 UT. The optical flux
densities at the wavelengths of $UBVRIHK$ bands, corrected for the
Galactic absorption, have been used without subtracting any host galaxy
contribution, because it was negligible at the epoch we selected
(cfr.~\cite{ref:fru}). When needed, we rescaled the data to the
corresponding reference date using the optical power--law decay with index
$\alpha_{\rm opt} = 1.10$~\cite{ref:kul}. We rescaled the flux of the
broadband X--ray spectrum (0.6--60 keV) using the power law decay measured
from 2--10 keV data ($\alpha_{\rm X} = 1.46$). Since the optical and
X--ray light curves showed different temporal decays, we independently
fitted with a power--law the optical/NIR spectrum and we obtained a
spectral index value of $\beta_{\rm opt} = 0.60\pm0.04$, flatter than the
X--ray one ($\beta_{\rm X} = 0.94\pm0.07$ at $1\sigma$). The presence of a
spectral turnover between optical and X--ray bands could be associated
with the change of spectral slope at a frequency which we identified with
the cooling frequency $\nu_c$ in the framework of the synchrotron fireball
model~\cite{ref:sar}. Assuming a negligible host absorption and using the
optical/NIR and X--ray slopes, we obtained for $\nu_c$ the value $1.1
\times 10^{17}$ Hz, corresponding to 0.5 keV.



\end{document}